\documentclass[aip,reprint]{revtex4-1}
\usepackage{dcolumn}% Align table columns on decimal point
\usepackage{bm}% bold math\begin{document}

\usepackage{epsfig}
\usepackage{wrapfig}

\newcommand{\apj}{ApJ}
\newcommand{\apjl}{ApJL}
\newcommand{\apjs}{ApJSupp}
\newcommand{\pasp}{PASP}
\newcommand{\mnras}{MNRAS}
\newcommand{\aj}{AJ}
\newcommand{\nat}{NATURE}
\newcommand{\aap}{A\&A}

\newcommand{\araa}{Ann. Rev. Ast. Ap.}
\newcommand{\aapr}{Ast. Ap. Rev.}

\begin{document}
\title{The Accretion of Solar Material onto White Dwarfs: 
No Mixing with Core Material Implies that the Mass of the White Dwarf is Increasing} 




\author{Sumner Starrfield}
\email{starrfield@asu.edu}

%\homepage[]{http://starrfield.asu.edu}

\affiliation{School of Earth and Space Exploration, \\ Arizona State University,\\
 P. O. Box 871404, Tempe, AZ 85287-1404, USA}

%\date{\today}

\begin{abstract}
Cataclysmic Variables (CVs) are close binary star systems with one component a white dwarf (WD) and the other a larger cooler star that fills its Roche Lobe.  The cooler star is losing mass through the inner Lagrangian
point of the binary and some unknown fraction of this material is accreted by the WD.  One consequence of the WDs accreting material, is the possibility that they are growing in mass and will eventually reach the Chandrasekhar Limit.  This evolution could result in a Supernova Ia (SN Ia) explosion and is designated the Single Degenerate Progenitor (SD) scenario.  One problem with the single degenerate scenario is that it is generally assumed that the accreting material mixes with WD core material at some time during the accretion phase of evolution and, since the typical WD has a carbon-oxygen (CO) core, the mixing results in large amounts of carbon and oxygen being brought up into the accreted layers.  The presence of enriched carbon causes enhanced nuclear fusion and a Classical Nova (CN)explosion.  Both observations and theoretical studies of these explosions imply that more mass is ejected than is accreted, and that the process repeats.  Thus, the WD in a Classical Nova system is decreasing in mass and cannot be a SN Ia progenitor.  However, the composition in the nuclear burning region is important and, in new calculations reported here, the consequences to the WD of {\it no} mixing of accreted material with core material have been investigated and it is assumed that the material involved in the explosion has only a Solar composition.  WDs with a large range in initial masses and mass accretion rates have been evolved.  I find that once sufficient material has been accreted, nuclear burning occurs in all evolutionary sequences and continues until a thermonuclear runaway (TNR) occurs and the WD either ejects a small amount of material or its radius grows to about $10^{12}$ cm and the calculations are stopped.  In all cases where mass ejection occurs, the mass of the ejecta is far less than the mass of the accreted material.  Therefore, all the WDs are growing in mass.  It is also found that the accretion time to explosion can be sufficiently short for a 1.0M$_\odot$ WD that a recurrent nova explosion can occur on a WD that is lower than typically assumed for the WDs in these systems.  Finally, the predicted surface temperatures when the WD is near the peak of the explosion imply that only the most massive WDs will be significant X-ray emitters at this time.  
\end{abstract}

%\pacs{}% insert suggested PACS numbers in braces on next line

\maketitle %\maketitle must follow title, authors, abstract and \pacs

\section{Introduction}

The two major suggestions for the objects that explode as a Supernova of Type Ia (SN Ia) are 
either the single degenerate (SD) or the double degenerate (DD) scenario.  In 
the standard paradigm SD scenario, it is proposed that 
a white dwarf (WD) in a close binary system accretes material from its companion and grows to the 
Chandrasekhar Limit.   As it nears the Limit, an explosion is initiated 
in the core.   In contrast, the double degenerate scenario (DD) 
requires the merger or collision of two WDs to produce the observed explosion.   
While for many years the SD scenario was the more prominent, a number of concerns led to major
efforts to better understand the DD scenario, in spite of the fact that the SD scenario is
capable of explaining most of the observed properties of the SN Ia explosions via the delayed detonation
hypothesis  \citep[and references therein]{khokhlov_1991_aa, kasen_2009_aa, woosley_kasen_11_a,
  howell_2009_ab}.   Reviews of the various proposals for SN Ia
progenitors \citep{branch_1995_aa}, producing a SN Ia, and the
implications of their explosions can be found in
\citet{hillebrandt_2000_aa},
\citet{leibundgut_2000_aa,leibundgut_2001_aa}, \citet{nomoto_2003_aa},
and \citet{howell_2011_aa}.

New evidence in favor of continuing the studies of the SD scenario come
from the observations of SN 2011fe in M101.  They show that 
the exploding star was likely a carbon-oxygen (CO) WD
\citep{pnugent11} with a companion that was probably on or near the main sequence
 \citep{weidongli_2011_aa,bloom_2012_aa}.  However, radio \citep{chomiuk_2012_aa} and optical \citep{bloom_2012_aa} observations
may have ruled out many types of Cataclysmic Variables (CVs) although
\citet{dilday_2012_aa} claim that PTF 11kx was a SN Ia that exploded
in a Symbiotic Nova system.  In addition, while \citet[]{schaeferpag_2012_aa} find no
star (to stringent but not impossible limits) at the ``center'' of a SN Ia remnant in the LMC, 
\citet[]{edwardspag_2012_aa} find a large number of stars near the ``center'' of a second LMC SN Ia remnant.  Then
\citet[]{schaeferpag_2012_aa} claim that they have ruled out the SD scenario (and \citet[]{edwardspag_2012_aa} claim that they do not),  but it is likely that there are either multiple SN Ia channels or the remnant secondary in the \citet[]{schaeferpag_2012_aa} study was fainter than their detection limit.   Nevertheless, the existence
of ``Super-Chandra'' SN Ias suggests that DD mergers are required for
these explosions.  The conclusion from these studies is that
there are multiple SN Ia channels and that each of them must be investigated.

Further support for the SD scenario, comes from observations of V445
Pup (Nova 2000) which imply that it was a helium nova (helium
accretion onto a WD) since there were no signs of hydrogen in the
spectrum at any time during the outburst, but there were strong lines
of carbon, helium, and other elements \citep[]{woudt_2005_aa,
woudt_2009_aa}.  Because it was extremely luminous before the
outburst, the secondary is thought to be a hydrogen deficient carbon
star \citep[]{woudt_2009_aa}.  Since one of the defining
characteristics of a SN Ia explosion is the absence of hydrogen or
helium in the spectrum at any time during the outburst or decline, the
existence of V445 Pup implies that mass transferring binaries exist in
which hydrogen is absent at the time of the explosion and most of the
helium is converted to carbon during the Classical Nova phase of evolution.

In this paper, therefore, I report on recent calculations that  explore the SD scenario which is based on the suggestion of \citet{whelan_iben_73} that the outburst occurs in a close binary system that contains a WD and another star.  
Since the WD is accreting material from a secondary, virtually every type of close binary has been suggested as a SN Ia progenitor.  Therefore, I investigate the evolution with accretion for a broad range in initial WD mass and mass accretion rate and follow the simulations with two different hydrodynamic computer codes. 

In the next section I discuss perceived problems with the SD scenario.  I follow that with a section that describes the two computer codes that I have used.  I follow that with the most important results from each code and end with a summary and discussion.

\section{Problems with the Single Degenerate Scenario}

Although, as noted above, the SD scenario can result in light curves and other explosion properties that resemble those of SN Ias, there are significant perceived problems with any of the suggestions for what the progenitors might actually be.  In fact, while virtually every type of close binary, or not so close binary, involving a WD has been suggested as a progenitor, a major problem is that there is no hydrogen or helium observed in the explosion.  Nevertheless, in virtually every observed binary that contains a secondary transferring material onto the WD, the material is hydrogen rich (except for V445 Pup as noted above).  The presence of hydrogen suggests either that these systems are not SN Ia progenitors or that the hydrogen and helium is lost from the system prior to the SN Ia explosion.   Moreover, many of the suggested classes of binaries are losing mass at prodigious rates into the local ISM. That material should still be nearby when the system explodes and it would then appear in the spectrum at some time after the outburst.  In fact, there are a few SNe Ia where there are narrow lines of hydrogen in the spectrum that indicate circum-binary or ISM material \citep{maoz_araa_13}.  In addition, there was sufficient hydrogen in the spectrum of  PTF 11kx that  \citet{dilday_2012_aa} claimed that it was a SN Ia that exploded in a Symbiotic Nova system.  

Another problem is that it is commonly assumed that only a very narrow range in mass accretion rate (\.M: M$_\odot$yr$^{-1}$) allows the mass of the WD to grow as a result of continued accretion.  The basis of this assumption is the work of \citet{fujimoto_1982_aa, fujimoto_1982_ab}.   An updated plot of his results can be found as Figure 5 in \citet{Kahabka_1997_aa} and I do not reproduce it here.  This plot has 3 regions on it.  For the lowest mass accretion rates, at all WD masses,  it is predicted that accretion results in hydrogen flashes that are predicted to resemble those of Classical Novae \citep{starrfield_2012_basi}.  Further, the results of a large number of observational studies, in combination with the theoretical predictions, of the amount of Classical Nova ejecta, imply that more mass is ejected than accreted \citep{gehrz_1998_aa, starrfield_2012_basi}.   In support of this assumption, hydrodynamic calculations of this process show that the accreted material must mix with core material, in order to produce a fast nova outburst, and then the explosion ejects both core and accreted material reducing the mass of the WD as discussed in \citet[(and references therein)]{gehrz_1998_aa}.  Therefore, at low \.M, if these predictions are correct, the systems cannot be SN Ia progenitors.

For the highest mass accretion rates in this plot, the results of \citet{fujimoto_1982_aa, fujimoto_1982_ab} imply that the radius of the WD rapidly expands to red giant dimensions, accretion is halted, and any further evolution must await the collapse of the extended layers.   There is, however, a third regime identified by \citet{fujimoto_1982_aa, fujimoto_1982_ab}, intermediate between these two, where the material is predicted to burn steadily at the rate it is accreted.  The central \.M of this region is nominally $\sim 3 \times 10^{-7}$M$_\odot$ yr$^{-1}$ but it does have a slight variation with WD mass.   The implication, therefore, is that only a narrow range of mass accretion rates results in a steady growth in mass of the WD. The observations of CVs and other systems with accreting WDs, however, show that they are accreting at rates that are not within the steady burning regime which suggests that the WD cannot be growing in mass.   

Those systems that are proposed to be accreting at the steady burning rate are thereby evolving to higher WD mass but, by some unknown mechanism, the mass transfer in the binary system is stuck in this mass accretion range.   The Super Soft Binary X-ray Sources (SSS) in the LMC are predicted to be in the steady burning regime \citep{vandenheuvel_1992_aa}.   Unfortunately, there are insufficient numbers of known SSS systems, thought to be accreting at these high rates, for them to be SN Ia progenitors. 

However, the calculations reported in \citep{fujimoto_1982_aa, fujimoto_1982_ab} on which this plot is based assume a steady state solution and imply that the only parameters that affect the evolution of an accreting WD are its mass and \.M.  His calculations  do not take into account 
the chemical composition of the accreting material, the chemical composition of the underlying WD, if mixing of 
accreted material with core material has taken place, or the thermal structure of the underlying WD. 
Moreover, they do not take into account the effects of previous (or continuing) outbursts on the thermal and compositional structure of the WD.  It is well known that all these parameters affect the evolution of the 
WD \citep[]{yaron_2005_aa,starrfield_2008_cn}.  In the next sections I report on two different studies of the accretion of Solar material onto WDs and show that the results of  \citet{fujimoto_1982_aa, fujimoto_1982_ab} are incomplete.

\section{The NOVA and Mesa Codes}

I report here on calculations done with two one-dimensional (1-D) hydrodynamic computer codes (NOVA and MESA) to study the accretion of only {\it Solar} composition material \citep{lodders_2003_aa} onto WD masses of 0.4M$_\odot$, 0.7M$_\odot$, 1.0M$_\odot$, 1.25M$_\odot$, and 1.35M$_\odot$.  I use two initial WD luminosities  ($4 \times 10^{-3}$ L$_\odot$ and $10^{-2}$L$_\odot$) and seven mass accretion rates ranging from $2 \times 10^{-11}$M$_\odot$ yr$^{-1}$ to $2 \times 10^{-6}$M$_\odot$ yr$^{-1}$.   In order to ensure that I used an \.M that overlapped with the steady burning regime of \citet{fujimoto_1982_aa, fujimoto_1982_ab}, I  added one study, at all WD masses, with an accretion rate of $3 \times 10^{-7}$M$_\odot$yr$^{-1}$.  I used an updated version of NOVA 
\citep[and references therein]{starrfieldpep09} that includes a nuclear
reaction network  that has 187 nuclei up to $^{64}$Ge.  The nuclear reaction rate library is described in  \citep{starrfieldpep09} and NOVA also includes the latest microphysics (equations of state, opacities, and electron conduction) and a new algorithm to treat mixing-length convection \citep{arnett_2010_aa}.  The simulations reported in this paper were done with 150 mass zones and with the surface zone mass less than $\sim 10^{-9}$M$_\odot$.  A few sequences were evolved with up to 395 mass zones and smaller surface zone masses in order to check the convergence of the results.  These latter changes had only small effects on the results.  However, NOVA can only follow the ``first'' outburst on a WD and it is not possible to determine if succeeding outbursts will change the
results.  In addition, it is necessary to follow multiple outbursts to determine the secular evolution of the accreting WD.

\begin{figure*}[htb!]
%\vspace{-13mm}
\center
\includegraphics[scale=0.6]{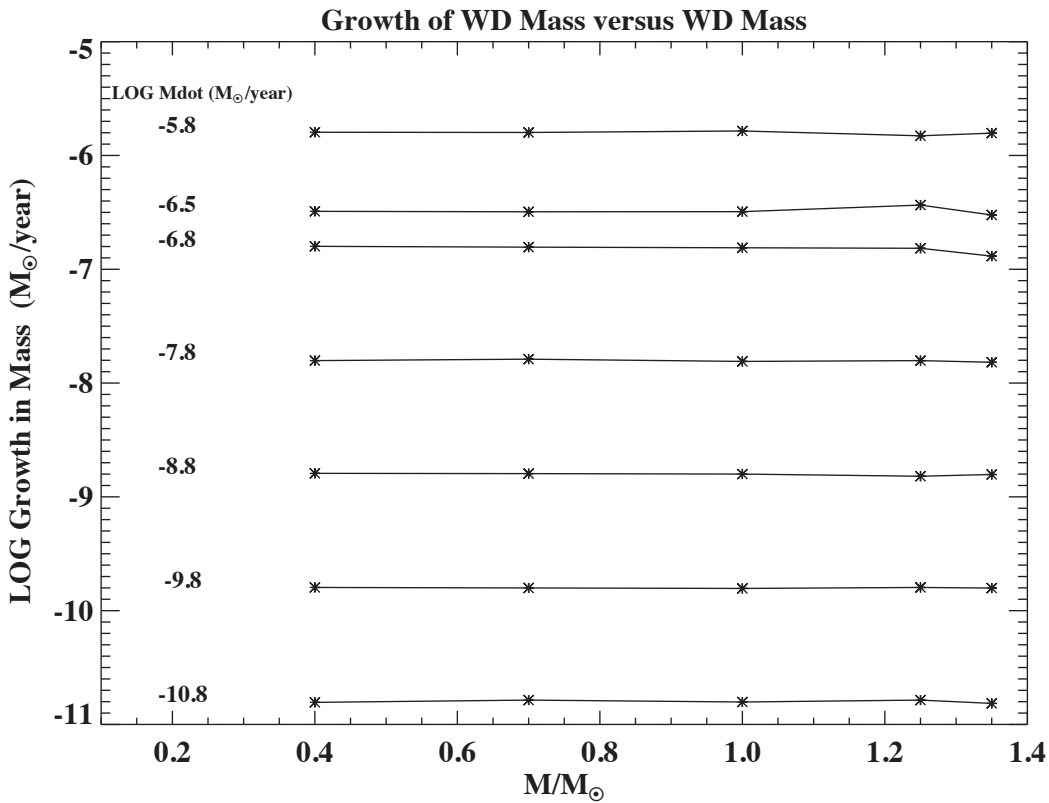}
%\vspace{-20mm}
\caption
{This plot shows the difference between the mass accreted and the mass lost for each of the sequences that we 
evolved in this study.  Since each point represents  the two initial luminosities that we used, there are 70 sequences shown here.  All sequences exhibited a TNR.  In no case did steady burning occur.  
 We display the growth in mass  (in units of M$_\odot$yr$^{-1}$)
 as a function of WD mass for each of our sequences.  Each point is the amount of accreted (less ejected) mass divided by the time to 
 reach the TNR for the given simulation.  The column of numbers on the left side of the plot is the Log of the mass accretion rate for all
 the points connected by the solid line.}
\end{figure*}

Therefore, a new stellar evolution code, MESA, was used because it is
capable of following multiple outbursts on an accreting WD.  It
solves the 1D fully coupled structure and composition equations
governing stellar evolution. It is based on an implicit finite
difference scheme with adaptive mesh refinement and sophisticated
time step controls; state-of-the-art modules provide equation of state,
opacity, nuclear reaction rates, element diffusion, boundary
conditions, and changes to the mass of the star
\citep{paxton_2011_aa, paxton_2013_aa}.  MESA has also been extended to include
new convection algorithms, oscillations, and rotation \citep{paxton_2013_aa}.

\begin{figure*}[htb!]
%\vspace{-15mm}
\center
\includegraphics[scale=0.6]{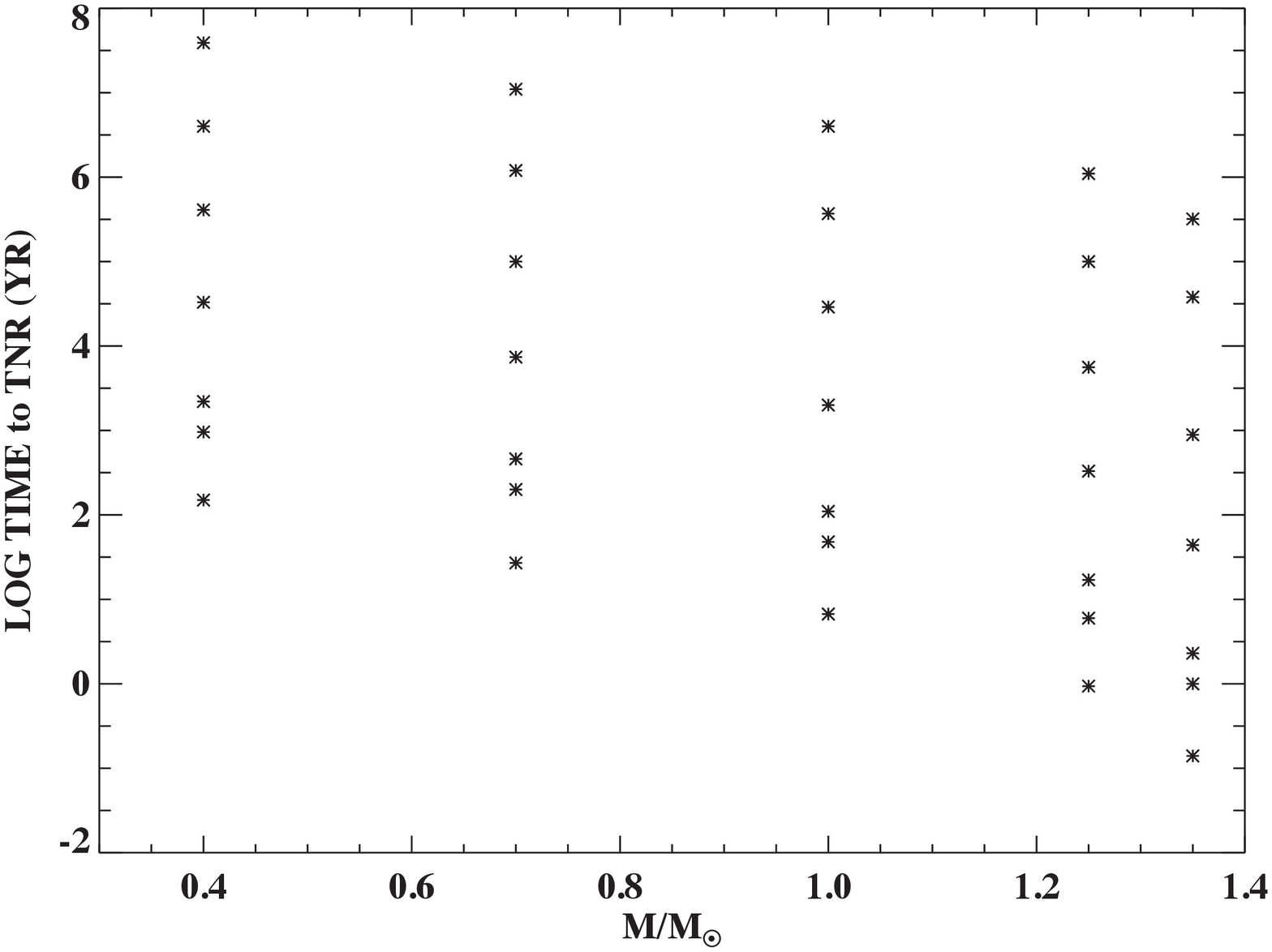}
%\vspace{-10mm}
\caption {This is a plot of the accretion time to the TNR as a function of WD mass.  Each of the data
points is for a different \.M and the value of \.M increases downward for each WD mass.  The accretion time,
 for a given \.M decreases with WD mass because it takes less mass to initiate the TNR as the WD mass
 increases.}
%\vspace{-2mm}
\end{figure*}

\begin{figure*}[htb!]
\center
%\vspace{-12mm}
\includegraphics[scale=0.6]{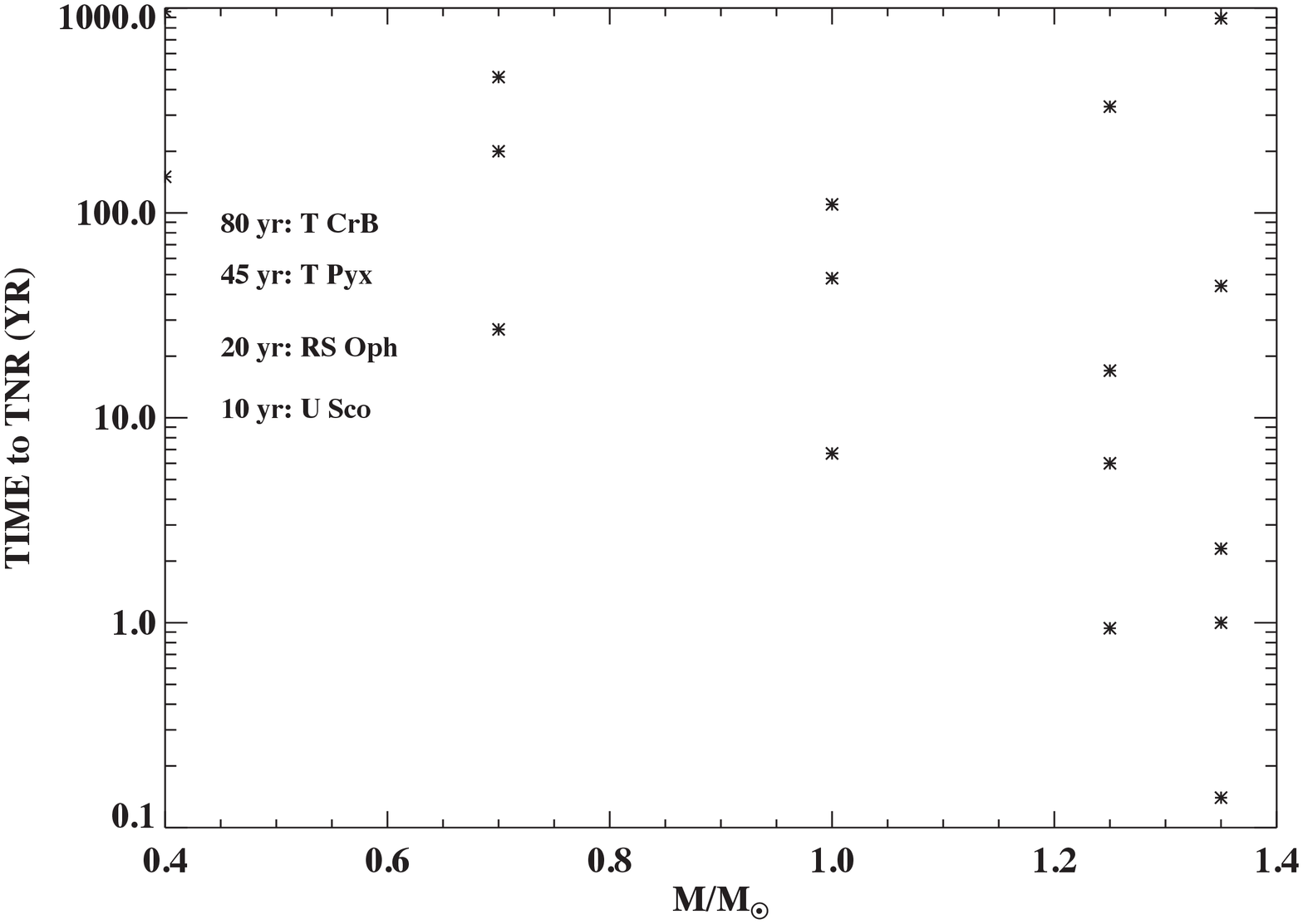}
%\vspace{-10mm}
\caption
{This is the same plot as Figure 2 but here we concentrate on the lower right corner and add 
 approximate recurrence times for the best known RNe.  The location of each RN indicates
 its approximate recurrence time.  This plot shows that not only is it possible for
 RNe outbursts to occur on low mass WDs but they can also occur for a broad range of \.M on higher mass WDs.}
\end{figure*}

\section{Results}
\subsection{Simulations with NOVA}

For each WD mass and \.M, an initial luminosity was chosen and it was assumed that no material had been
accreted prior to the beginning of the evolution.  The accretion rate was kept constant for each of the 70 
simulations.  The accreting material was allowed to move through the Lagrangian mesh as described by
\citet{kuttersparks_1987_aa} until it reached the temperatures at which nuclear burning was
initiated \citep{starrfieldpep09}.  No mixing of accreted material with core material 
was allowed.   Once nuclear burning was proceeding at a sufficient rate and convection had begun
just above the core-accreted matter interface (peak temperature $\sim 2\  \rm to\  3 \times 10^7$K),  the accretion algorithm was changed to that described in \citet{kuttersparks_1980_aa}.  This change was done to ensure that the core-accreted matter interface occurred on a Lagrangian boundary of the mesh.  Accretion was then continued until the convective region had reached about half-way from the core-accreted matter interface to the surface.  By this time the peak temperature at the core accreted-matter interface was $\sim 5\  \rm to\  6 \times 10^7$K.    Accretion was then ended, a rapid increase in temperature to peak nuclear burning occurred (thermonuclear runway: TNR), and the resulting evolution was followed through peak conditions and expansion of the surface layers to $> 10^{12}$cm.    Only a few of these simulations ejected any material and the amount ejected was far less than the amount accreted.  The difference between these results and those with enriched nuclei in the nuclear burning region is that, with fewer catalytic nuclei present, there is insufficient energy produced at the peak and just after the peak of the TNR to drive a significant amount of material out of the potential well of the WD.

In all simulations, the evolution was ended when the radius of the surface layers of the WD had grown to $\sim 10^{12}$cm.   The amount of material accreted minus that ejected (the material velocity exceeded the escape velocity at this radius), if any, was tabulated and is shown in Figure 1.  This figure shows the results for all 70 simulations (each data point represents two initial luminosities).   It gives the mass accreted, minus mass lost, as a function of WD mass for each simulation.  The value of \.M is given to the left of each set of data points connected by a line.  This plot shows that all WDs are growing in mass as a result of the
accretion of Solar material.  

However, there is material at $\sim 10^{12}$cm that has not reached escape
velocity.   This radius exceeds the Roche Lobe radius of most observed CVs and I assume that it is undergoing a common envelope phase of evolution so that a small 
additional amount of material will be ejected by the secondary star orbiting within the extended layers of the WD 
as discussed in detail in \citet{macdonald_1986_aa}.  This additional amount is not included in Figure 1 because, in all cases, it was only a few percent of the accreted material.

Since the implication of steady burning as described by
\citet{fujimoto_1982_aa, fujimoto_1982_ab} is that the material burns to helium at exactly the
rate at which it is being accreted,  in none of these simulations did canonical steady burning occur.  In contrast, a TNR occurred, the temperature in the nuclear burning region rose to exceed the Fermi temperature, and the accreted material expanded to large radii.  In fact, these fully time-dependent calculations show that these sequences exhibited the \citet{schwarzschild_1965_aa} hydrogen thin shell instability and the existence of this instability implies that steady burning cannot occur.   An expanded study of the stability of thin shells can be found in \citet{yoon_2004_aa} who investigated the accretion of hydrogen-rich material onto WDs.  They established regions of steady and unsteady burning and using their results, I find that the simulations reported here are initially in a stable region (see their Figures  8 and 11) but with continued accretion evolve into instability.        

\begin{figure*}[htb!]
%\begin{figure}[]
\center
%\vspace{-12mm}
\includegraphics[scale=0.6]{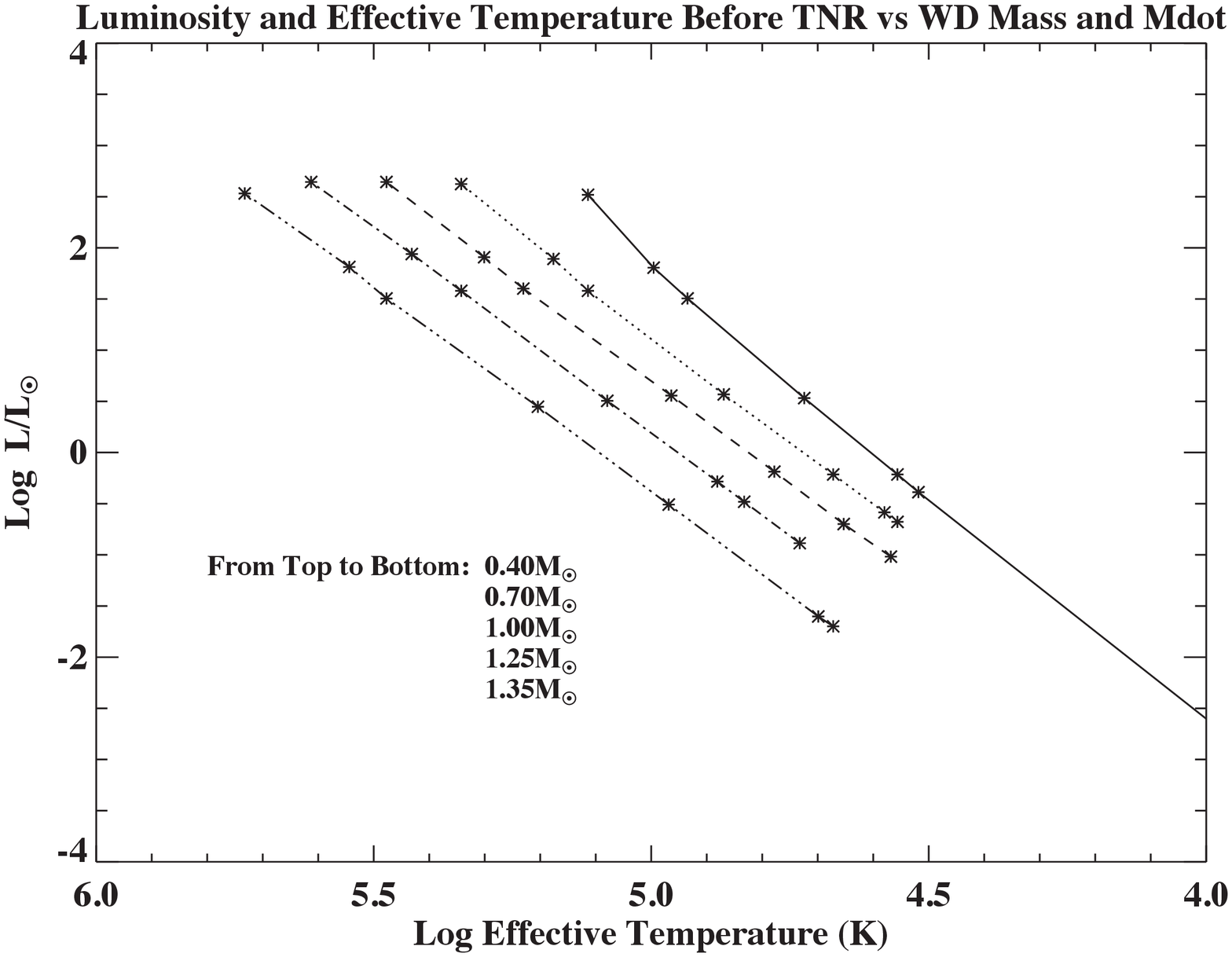}
%\vspace{-10mm}
\caption
{This plot shows the luminosity and effective temperature in the HR diagram for the simulations around the time
just before the final rise to the peak of the TNR.  The differences in luminosity and effective temperature along each line is caused by a difference in \.M.  \.M increases from right to left along each line.  Most of these sequences, especially the ones occurring on the lower mass WDs, would not be detected by the current low energy detectors on the X-ray satellites (see text).  The WD mass is labeled on the figure and it goes from 0.4M$_\odot$ (straight line) to
1.35M$_\odot$ (the dash dot dot dot line).}
\end{figure*}

It is also the case that the low mass WDs did not eject any mass (ignoring the common envelope phase) while the high mass WDs eject only a small fraction of their accreted material (less than 10\%).  Therefore, the WDs are growing in mass as a result of the accretion of Solar material and mixing of accreted with core material is not allowed.  (This is not the case for Classical Novae which show sufficient core and accreted material in their ejecta that the WD must be losing mass as a result of the outburst.)   We identify the systems, where I predict that the WD is growing in mass, with those Cataclysmic Variables (Dwarf Novae, Recurrent Novae, Symbiotic Novae, ...) that show no core material either on the surface of the WD or in their ejecta.   These results could explain those of \citet{Zorotovic_2011_aa} who report that the WDs in CVs are growing in
mass.  In addition, the best studied Dwarf Novae have WD masses larger
than the canonical value of $\sim$0.6M$_\odot$ for single WDs.  These are U Gem: 1.2M$_\odot$
\citep{Echevarria_2007_aa}, SS Cyg: 0.8M$_\odot$
\citep{Sion_2010_SSCyg_aa}, IP Peg: 1.16M$_\odot$
\citep{copperwheat_2010_aa}, and Z Cam: 0.99M$_\odot$
\citep{Shafter_1983_aa}.

I also show the accretion time to TNR for all the sequences (Figure 2).  As is well known \citep{yaron_2005_aa, starrfield_2008_cn}, as the WD mass increases, the accretion time decreases for the same \.M.  In addition, as \.M increases, the time to TNR decreases.  This behavior occurs because higher mass WDs have a smaller radius and, thereby, a higher gravitational potential energy, so that they are able to initiate the TNR with a smaller amount of accreted mass.  In Figure 3, I concentrate on the lower right corner of Figure 2 and add approximate recurrence times for the best known recurrent novae (RNe). Although it is typically claimed that only the most massive WDs can exhibit recurrence times short enough to agree with those of the listed (and best known) RNe, this plot shows that this is not the case.  It is possible for RNe to occur on WDs with masses as low as 0.7M$_\odot$.   Therefore, basing the WD ``masses'' of  RNe on short recurrence times is incorrect.  This plot also shows that it is possible for a RN outburst to occur on a high mass WD for an extremely broad range of \.M.  I note here that there is a RN in M31 that has undergone outbursts about once a year for 7 years and this behavior can be explained by accretion at high rates onto a WD with a mass exceeding $\sim 1.3$M$_\odot$.

Another important result arises from the claims that there are insufficient CV systems identified
in various X-ray searches for them to be SD SN Ia progenitors \citep[e.g.,][]{gilfanov_2010_aa}.  I investigated
this question by tabulating the effective temperatures and luminosities of the simulations both during the evolution to
the TNR (Figure 4) and at the peak of the TNR (Figure 5).  

To better understand the implications of these plots, I refer to the evolution of RS Oph in X-rays \citep{osborne_2011_aa}.   \citet{osborne_2011_aa} analyzed the Swift X-ray light curve of 
RS Oph and found that this RN did not start to become a Super Soft Source (emit a large number of X-rays with energies below about 0.5 keV) until about day 26 of the outburst.  They interpreted this behavior as the consequences of nuclear burning on the surface of
the WD causing its effective temperature to gradually increase until it became sufficiently hot for the emission
to be detected by the Swift satellite.  Using a calculation from \citet{bath_1989_aa}, they estimated that it was not detected by Swift until the temperature of the nuclear burning WD had reached to $\sim$400,000K .  This high a temperature agrees with analyses of X-ray grating spectra obtained at about the same time \citep{ness_2007_ab}.   Given the poor low energy response of the CCD detectors on Swift and other X-ray satellites,  a WD evolving to a TNR must exceed an effective
temperature of at least 400,000K before it can be detected in X-rays.   Figure 4 shows that only the
most massive WDs, accreting at the highest mass accretion rates, will be detected as SSSs in
X-rays.  However, these systems also have the shortest ``duty'' cycles and could be missed
on their evolution to explosion.  So, their non-detection is not surprising.

\begin{figure*}[htb!]
%\begin{figure}[]
\center
%\vspace{-12mm}
\includegraphics[scale=0.6]{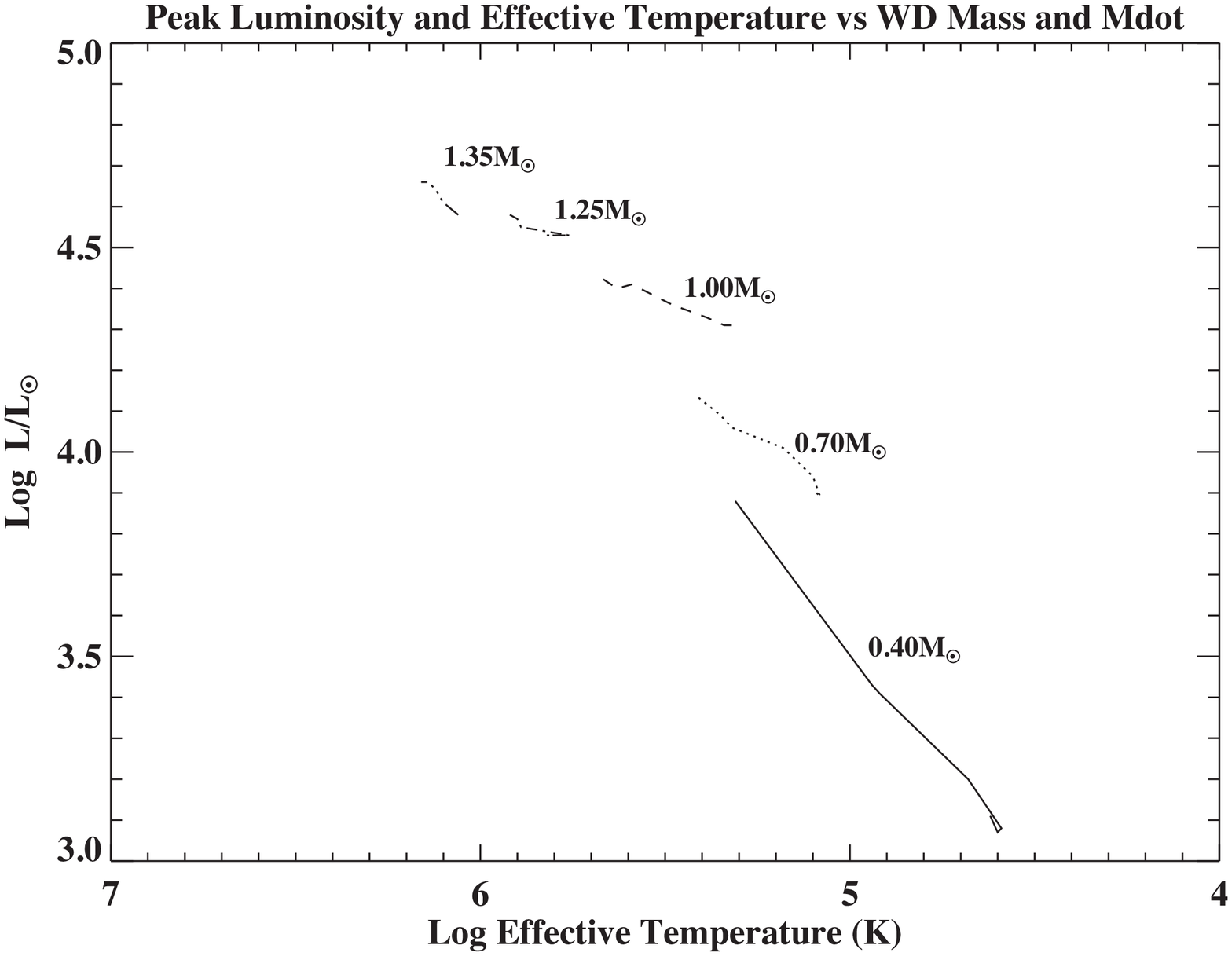} 
%\vspace{-10mm}
\caption{This is the same plot as in Figure 4 but for the peak conditions during the TNR. 
All but those occurring on the lowest mass WDs would be detected by current X-ray satellites.  However, those on the
highest mass WDs, and thus closest to the Chandrasekhar Limit, would be detected for the shortest amount
of time. }
\end{figure*}

Figure 5 shows these systems at their peak effective temperature in the HR diagram during the explosive phase.
The sequences that are the hottest and most
luminous are again those with the highest mass accretion rate at each WD mass.  They are also the sequences that
have accreted the least amount of material and, therefore, have ejected the least amount of material.  They will
be ``bright'' in X-rays for the shortest amount of time.  The results shown in Figure 5 imply that only high mass WDs will be detected in X-rays at maximum and that if a CN is detected in soft X-rays during the outburst it occurred on a massive WD.

\subsection{Simulations with  MESA}

For the studies with MESA, in which it is possible to follow multiple outbursts, I only used a subset
of the WD masses reported on in the last subsection: 0.7M$_\odot$, 1.0M$_\odot$ and 1.35M$_\odot$.
All WDs consisted of initially bare CO cores (C = 0.357, O = 0.619) prior to the beginning of accretion.
The initial models had a Solar luminosity.  The mass accretion rates were chosen to be
$1.6 \times 10^{-10}$M$_\odot$yr$^{-1}$, $1.6 \times 10^{-9}$M$_\odot$yr$^{-1}$, 
$1.6 \times 10^{-8}$M$_\odot$yr$^{-1}$, and $1.6 \times 10^{-7}$M$_\odot$yr$^{-1}$
Other accretion rates were used in order to separate different regimes of behavior when needed. 
The material being accreted was also a Solar mixture as in the studies with NOVA \citep{lodders_2003_aa}.  
All simulations were run for either many nova cycles or until long-term behavior became evident.

\begin{figure*} [htb!]
\center
\includegraphics[scale=0.60]{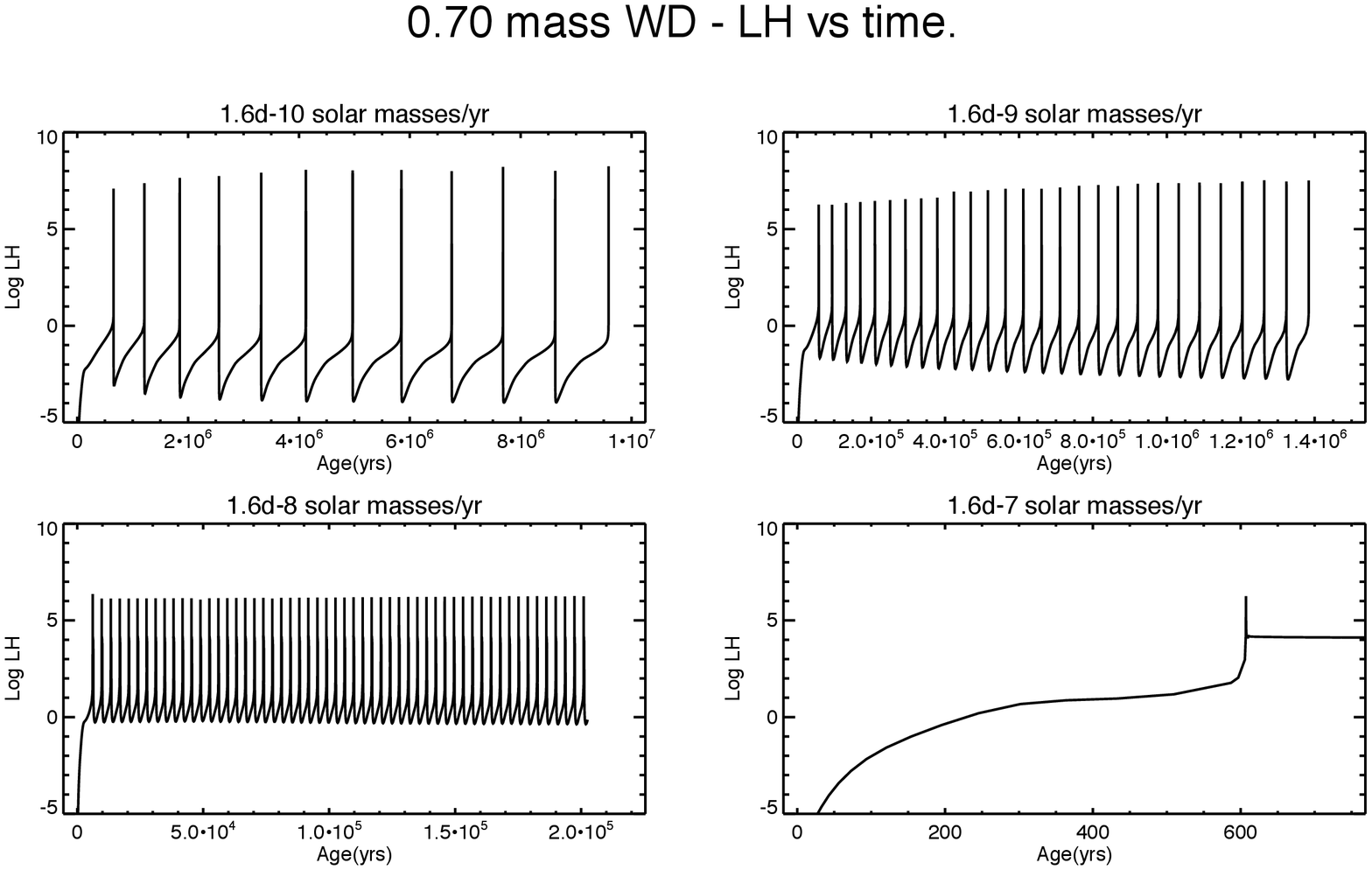}
\caption{This plot shows the log of the hydrogen luminosity as a function of time for accretion onto a
0.7M$_\odot$ WD.  There are 4 different mass accretion rates shown here and they are identified at the
top of each panel.  As the mass accretion rate increases, the time between outbursts decreases. The peak
luminosity, however, remains about the same because it depends on WD mass.  The sequence with the highest \.M (lower right hand side) accretes until a flash occurs and then grows rapidly to a radius of $10^{12}$cm and the evolution is ended.  The surface luminosity for this sequence exceeds the values measured for CVs}
\end{figure*}

Because MESA  can follow the long-term behavior of the accreting WD, through multiple TNR cycles,
the treatment of mass loss just after a TNR is important.  The simulations done with NOVA already show that, in all cases, either mass loss occurs or the radius grows to at least $10^{12}$cm.  
NOVA treats mass loss by following the velocities
of the material and their optical depth.  Once an expanding mass zone reaches escape velocity and has 
become optically thin, it is considered to have escaped.  But, because removing the material would effect the
numerical pressure on the inner zones, it is not actually removed.  The simulation is ended and the
amount of mass that has escaped is tabulated.   

This cannot be done in MESA since the escaping
zones must be removed in order to initiate a new accretion phase.  MESA does allow for 
different prescriptions of mass loss \citep{paxton_2011_aa, paxton_2013_aa}.  Here, a 
prescription based on the super-Eddington wind model of \citet{shavivnj_2002_aa} was used.
When a simulation reaches super-Eddington luminosities in the outer layers, the excess luminosity over Eddington determines the rate of mass loss in the WD gravitational potential.  A comparison of the \citet{shavivnj_2002_aa} model mass loss to that in NOVA shows that more mass is ejected by the \citet{shavivnj_2002_aa} model during the late stages of the flash.  This means that the MESA study is more conservative and that the WD mass grows more slowly than if the calculations had been done with just NOVA.  Nevertheless, the amount of mass lost during each flash depends on the method used to remove mass and the rate of growth in WD mass also varies according to the particular method.  Further work in this area is warranted.

In Figure 6, I show the logarithm of the hydrogen luminosity versus time for the 0.7M$_\odot$ evolutionary sequences accreting at four different rates: $1.6 \times 10^{-10}$M$_\odot$yr$^{-1}$, $1.6 \times 10^{-9}$M$_\odot$yr$^{-1}$, $1.6 \times 10^{-8}$M$_\odot$yr$^{-1}$, $1.6 \times 10^{-7}$M$_\odot$yr$^{-1}$.  While the peak luminosity is approximately the same for all four mass accretion rates, the time between outbursts decreases as \.M increases.  The sequence with the highest value of \.M undergoes only one outburst after which it steadily grows in mass.  The WD mass is also growing in the other cases.   The same 4 accretion rates have been applied to 1.0M$_\odot$ and 1.35M$_\odot$ WDs.  Because the mass of the WD is larger, it takes less accreted material to achieve a TNR and thus the time between outbursts decreases for the same \.M used in the 0.7M$_\odot$ studies.  

\begin{figure*} [htb!]
\center
\includegraphics[scale=0.60]{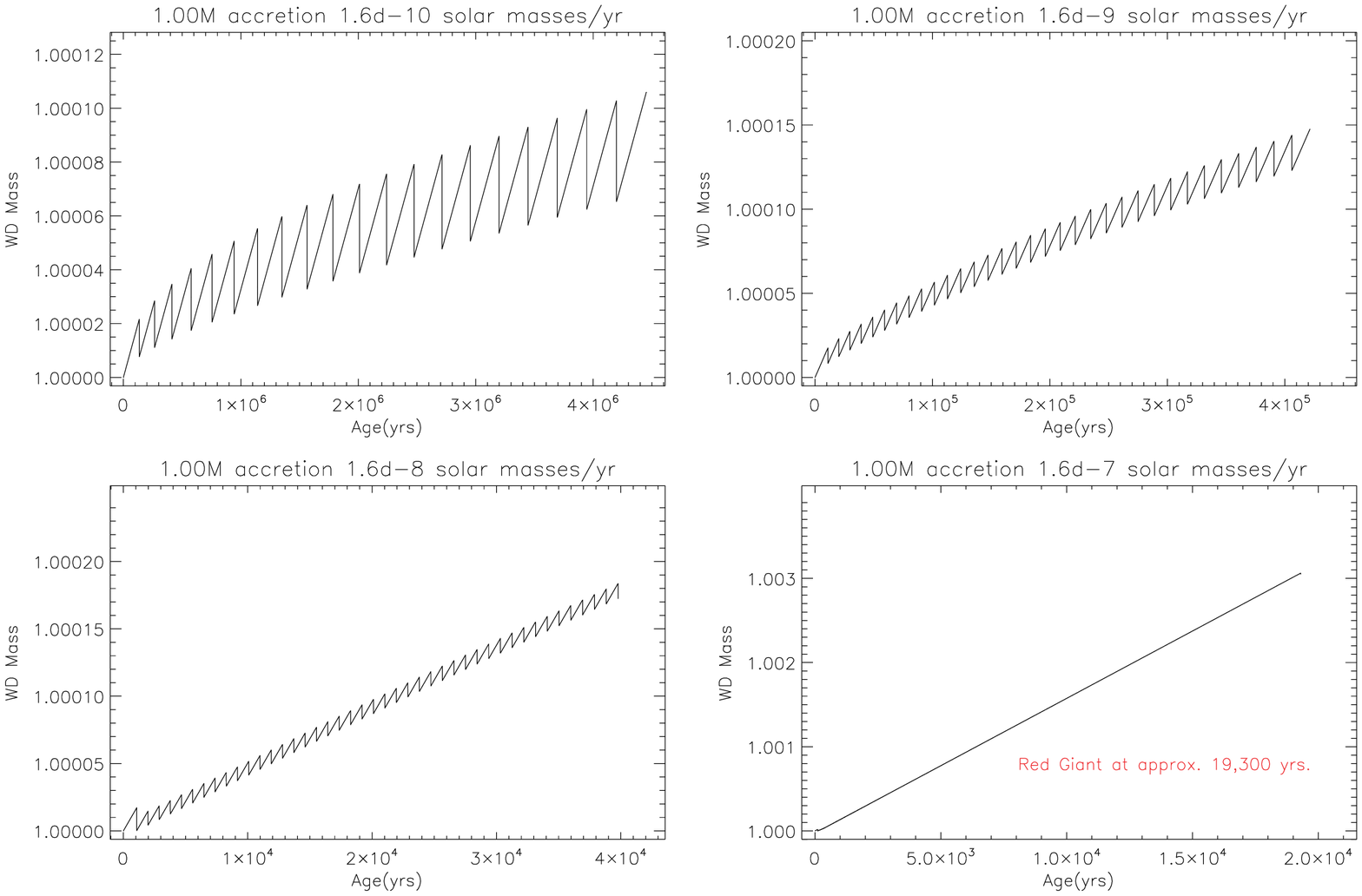}
\caption{This plot shows the WD mass as a function of time for a 1.0M$_\odot$ WD accreting at the rates listed
on top of each panel.  The jagged shape is caused by the mass growing as the accretion continues but then decreasing during each flash.  Nevertheless, the secular evolution is such that the WD is growing in mass at these
accretion rates.  The sequence with the highest \.M (lower right panel) accretes with small amplitude flashes for 19,300 yr before growing to red giant dimensions at which time the evolution is ended.  This is a higher \.M than is observed for CVs. }
\end{figure*}

Figure 7 shows the growth in mass for the 1.0M$_\odot$ WD accreting at the four different mass accretion rates. The value of \.M is listed on top of each panel.  The WD mass grows as it accretes and then decreases during the outburst as mass is lost via the prescription described above.  While the mass loss - mass gain curves show large amplitudes for the two lower mass accretion rates, the secular slope is upward.  The WD is gaining in mass.  The same happens at the two higher mass accretion rates but the amplitude for the $1.6 \times 10^{-7}$M$_\odot$yr$^{-1}$ simulation is less.  Unlike simulations at lower \.M, after 19,300 years of evolution at $1.6 \times 10^{-7}$M$_\odot$yr$^{-1}$ the sequence grows to red giant dimensions and the evolution is stopped.  Although not shown here, the mass gain for the 0.7M$_\odot$  sequences shows the same behavior:  large amplitude mass gain - mass loss cycles that show that the WD is gaining in mass.  In addition,  the highest mass accretion rate also grows to red giant dimensions.   

The behavior at the highest WD mass, 1.35M$_\odot$, is similar to the evolution at lower accretion rates but there is a transient ejection event for the first flash.  After the initial growth to the first flash, the simulation again (just as in the simulations at lower \.M) quickly settles into a recurring pattern of flashes in between which mass is accreted, lost during the flash, and then increases again in the next accretion phase.   The WD mass is growing with time at a rate of  $\sim 3.0 \times 10^{-8}$M$_\odot$yr$^{-1}$ for an accretion rate of  $1.6 \times 10^{-7}$M$_\odot$yr$^{-1}$.  This represents an efficiency (defined as the mass accreted minus the mass ejected divided by the mass accreted over a flash cycle) per cycle of approximately 20\%.   However, the 1.35M$_\odot$ sequences, at the highest accretion rates, exhibit a different behavior from the lower WD mass simulations.   After an initial hydrogen flash the sequence evolves into a steady-burning phase interrupted by regular helium flashes with a recurrence time of approximately 75 years.  During these helium flashes, about half the accreted mass is ejected from the WD.  Nevertheless, the WD continues to grow in mass at a rate of $2.6 \times 10^{-7}$M$_\odot$yr$^{-1}$ with an accretion efficiency of 41\%.

\begin{figure*} [htb!]
\center
\includegraphics[scale=0.80]{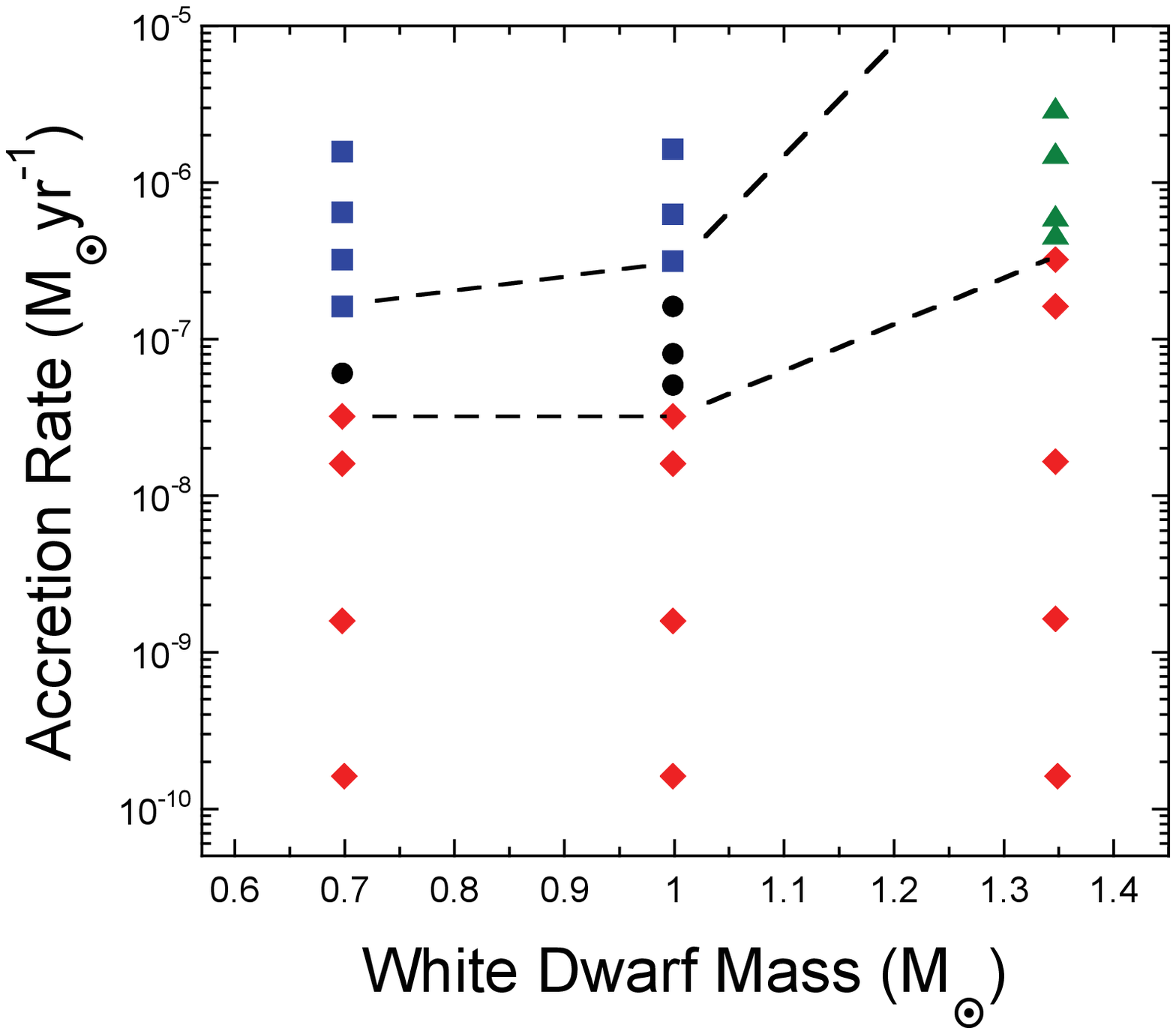}
\caption{Mass accretion rate versus WD mass plotted for the range of sequences
that we investigated. The symbols indicate sequences that become red
giants (blue squares), steady-burning followed by hydrogen flashes
(blue circles), steady-burning interrupted by helium flashes (green triangles)
and recurrent hydrogen flashes (red diamonds).  The steady-burning
region is found between the dashed lines although it does not exist at
1.35M$_{\odot}$.  In all cases where the properties of the evolutionary sequences resemble
those observed for CVs, the WD is growing in mass.  Since the typical observed \.M for CVs lies
below $\sim 10^{-8}$M$_\odot$yr$^{-1}$, the WD must be growing in mass for all WD masses.}
\end{figure*}

Finally in Figure 8, I summarize these calculations with a plot of WD mass versus \.M.  plotted for the range of sequences that were studied. The various symbols indicate sequences that became red
giants (blue squares), long phases of steady accretion followed by hydrogen flashes
(black circles), long phases of steady accretion interrupted by helium flashes (green triangles)
and recurrent hydrogen flashes (red diamonds).   All sequences below the lower dashed line
grow steadily in mass as they undergo repeated TNRs and mass accretion-mass loss cycles. 
This is in stark contrast to the
picture from  \citet{fujimoto_1982_aa, fujimoto_1982_ab} where this region
is filled with WDs accreting, experiencing Classical Nova outbursts, and then declining in
mass.   The region between the two dashed lines for WDs with masses less than 1.35M$_\odot$
are those where there are long periods of steady growth in mass followed by short episodes
of mass loss.  However TNRs occur for all these simulations.  The blue squares are the 
regions of highest mass accretion for lower mass WDs where the WD does grow to large
radii and the simulation is ended.  This behavior does not occur for the 1.35M$_\odot$ simulation
at the highest  \.M.  These simulations accrete until a helium flash occurs.  However, not all the
material is ejected so that the WD is still increasing in mass.   Nevertheless, the accretion plus 
nuclear burning luminosity for these simulations is so high that they would be easily observable.  
It is possible that this is the explanation for the existence of the Super Soft X-ray Binary Sources as
proposed by \citet{vandenheuvel_1992_aa} (see also \citet{Kahabka_1997_aa}).

\section{Summary and Conclusions}

I have studied the accretion of Solar material onto WDs with masses ranging from 0.4M$_\odot$ 
to 1.35M$_\odot$.  The seven mass accretion rates used in this work ranged from $2 \times 10^{-11}$M$_\odot$ yr$^{-1}$ to $2 \times 10^{-6}$M$_\odot$ yr$^{-1}$.  I also used two different hydrodynamic stellar evolution codes NOVA and MESA.  With NOVA I was able to study a broader range in WD mass and \.M but could only evolve the first outburst on the WD.  With MESA, I was able to follow a large number of outbursts and determine the secular evolution of the WD in response to mass accretion.  A TNR occurred for all 70 cases evolved with NOVA.  
In a few cases a small amount of mass was ejected but in most of these cases the surface layers of the WD just expanded to a radius of $\sim 10^{12}$cm and the evolution was ended.  In no case with NOVA did steady burning occur.  This result is in agreement with the work of \citet{schwarzschild_1965_aa} who first discovered the hydrogen thin shell instability in non-degenerate material.  A more recent study of accretion onto WDs can be found in \citet{yoon_2004_aa}.  Examining  their work, the calculations done with NOVA are initially in their stable regime but evolve into instability.  

I followed the NOVA simulations with a new set using MESA because this code can follow repeated outbursts on a WD and it was necessary to determine the secular evolution of the WD.   While the results with MESA appear similar to the plot shown in the work of  \citet{fujimoto_1982_aa, fujimoto_1982_ab} as given in 
\citet{Kahabka_1997_aa}, in fact there are large differences.  As shown in Figure 8,  all evolutionary sequences below the bottom dashed line are growing in mass.  They are not undergoing hydrogen flashes that eject more mass than is accreted.  These are the mass accretion rates determined for typical CV's so that if they are only accreting Solar material and there is no mixing of accreted with core material, the WDs in these systems are growing in mass.  The ``canonical'' steady burning regime (between the two dashed lines) delineates the region where long periods of accretion occur that are interrupted by hydrogen flashes.  Some of the flashes eject a significant amount of material but not as much as has been accreted.  For the lower mass WDs, this region lies below that identified in the Fujimoto version of this diagram.  At the highest mass WD, 1.35M$_\odot$, the long periods of accretion are broken by helium flashes that eject a large amount of material but, again, not as much as has been accreted. Therefore, WDs accreting at these rates are also growing in mass.  It is entirely possible that the Supersoft Sources first identified in the LMC are accreting in this regime and their WDs are growing in mass as has already been proposed by \citet{vandenheuvel_1992_aa} (see also \citet{Kahabka_1997_aa}).  

Finally, given the success of these calculations at growing the mass of the WD, it is appropriate to discuss the basic assumption that mixing of accreted with core material does not occur under all circumstances.  While there have been a number of sophisticated multidimensional simulations of mixing that lead to Classical Nova outbursts 
\citep{casanova_2011_aa, casanova_2011_ab}, because of CPU time limitations they have all been done for times shortly after convection has begun at the core-accreted material interface and are limited to a few 100 seconds of evolution time or less.  It is not possible to do multidimensional calculations from the beginning of accretion and follow them through the peak of the TNR.  Therefore, it is necessary to turn to the observations of both CVs (dwarf novae, AM Her variables, etc.) and Classical Novae and Recurrent novae.  In fact, while the observations of Classical Novae ejecta show sufficiently enriched CNONeMg elements that they must come from core material \citep{gehrz_1998_aa}, observations of CVs show little or no enrichment.  Since only a small number of CVs show ejected shells 
\citep{shara_2012_aa, shara_2012_ab}, this is an area that needs further work.  I end by referring back to earlier statements that the WDs in CVs appear to be growing in mass and, in addition, the WDs in the 4 nearest (and probably best studied) dwarf novae are more massive than the canonical mass of a single WD of $\sim 0.6$M$_\odot$.

The conclusions to this work are:

\begin{itemize}

\item Simulations of accretion of solar material onto WDs always produce a thermonuclear
runaway and ``steady burning'' does not occur.

\item Thermonuclear runaways on more massive WDs (than 0.4M$_\odot$) eject material but not much as 
compared to the amount accreted to initiate the runaway.

\item All WDs in CVs are growing in mass as a consequence of the accretion of Solar material.

\item The time to runaway is sufficiently short for accretion onto most of the WD masses that were
studied that Recurrent Novae could occur on a much broader range of WD mass than heretofore
believed. This is especially true for the newly discovered M31 RN that appears to be outbursting 
about once per year.

\item During most of the evolution time to the peak of the thermonuclear runaway the surface 
conditions of the WD (effective temperature and luminosity) are too low to be detected by the
currently orbiting low energy X-ray detectors.  Their non-discovery is not surprising. 

\end{itemize}

\acknowledgments

I acknowledge useful discussions with F. X. Timmes.  The MESA simulations were done by G. Newsham
who was supported by an NSF grant to ASU. 
I thank J. van Loon for his comments, as referee, which improved this paper.
Starrfield is happy to acknowledge partial support from U. S. National Science Foundation and NASA grants to ASU.

%\end{acknowledgements}

%\bibliographystyle{apj}
%\bibliographystyle{aipnum4-1}

%\bibliography{references_iliadis,starrfield_master,timmes_nasa}


\end{document}